\documentclass[aps,notitlepage,nofootinbib,twocolumn,longbibliography,10pt]{revtex4-1}

\newcommand{\beq}{\begin{eqnarray}}
\newcommand{\eeq}{\end{eqnarray}}
\newcommand{\beqq}{\begin{eqnarray*}}
\newcommand{\eeqq}{\end{eqnarray*}}

\newcommand{\be}{\begin{equation}}
\newcommand{\ee}{\end{equation}}
\newcommand{\barr}{\begin{array}}
\newcommand{\earr}{\end{array}}

\newcommand{\qtot}{Q_\textrm{tot}}
\newcommand{\ptot}{P_\textrm{tot}}
\newcommand{\rhored}{\rho_\textrm{red}}
\newcommand{\rhoblue}{\rho_\textrm{blue}}

\usepackage[colorlinks=true,citecolor=blue,linkcolor=blue]{hyperref}
\usepackage{graphicx}
\usepackage{dcolumn}
\usepackage{bm}
\usepackage{color}
\usepackage{tabularx}
\usepackage{comment}
\usepackage{float}
\usepackage{amsmath}
\usepackage{gensymb}
\usepackage{mathtools}
\usepackage{tikz-cd}
\usepackage{adjustbox}
\usepackage{braket}

\begin{document}

\begin{titlepage}

\widetext

\title{Hilbert space fragmentation produces an effective attraction between fractons}

\author{Xiaozhou Feng}
\author{Brian Skinner}
\affiliation{Department of Physics, The Ohio State University, Columbus,
Ohio 43202, USA}

\setcounter{equation}{0}
\setcounter{figure}{0}
\setcounter{table}{0}

\makeatletter
\renewcommand{\theequation}{S\arabic{equation}}
\renewcommand{\thefigure}{S\arabic{figure}}
\renewcommand{\thetable}{S\Roman{table}}
\renewcommand{\bibnumfmt}[1]{[S#1]}
\renewcommand{\citenumfont}[1]{S#1}

\date{\today}

\begin{abstract}
Fracton systems exhibit restricted mobility of their excitations due to the presence of higher-order conservation laws. Here we study the time evolution of a one-dimensional fracton system with charge and dipole moment conservation using a random unitary circuit description. Previous work has shown that when the random unitary operators act on four or more sites, an arbitrary initial state eventually thermalizes via a universal subdiffusive dynamics.  In contrast, a system evolving under three-site gates fails to thermalize due to strong ``fragmentation'' of the Hilbert space. Here we show that three-site gate dynamics causes a given initial state to evolve toward a highly nonthermal state on a time scale consistent with Brownian diffusion. Strikingly, the dynamics produces an effective attraction between isolated fractons or between a single fracton and the boundaries of the system, as in the Casimir effect of quantum electrodynamics. We show how this attraction can be understood by exact mapping to a simple classical statistical mechanics problem, which we solve exactly for the case of an initial state with either one or two fractons.


\end{abstract}


\maketitle

\draft

\vspace{2mm}

\end{titlepage}

\section{Introduction}
\label{sec:intro}

A fracton is a kind of excitation in certain quantum systems that exhibits reduced or fractionalized mobility, such that no local operator can move the fracton without producing additional excitations \cite{PhysRevLett.94.040402,BRAVYI2011839,Haah2011,Vijay_Haah_Fu_2015,Vijay_Haah_Fu_2016,Pretko_2017, gromov_classification_2019, Doshi2021} (see also Refs.~\onlinecite{doi:10.1146/annurev-conmatphys-031218-013604, pretko_review_2020} for reviews). Such reduced mobility can be framed in terms of higher-order conservation laws \cite{Pretko_2017, Ma_Hermele_Chen_2018,Bulmash_Barkeshli_2018}. A paradigmatic example, which we focus on in this paper, is that of a one-dimensional system with both charge and dipole moment conservation. 
In this setting, a single charge can only move in a given direction if a dipole is simultaneously created from the vacuum that points in the opposite direction \cite{Pretko_2017, pretko_gauge_principle_2018}. Such restricted mobility dynamics has led to intense recent interest in fracton systems as candidates for quantum many-body systems that fail to thermalize even at infinite time (e.g., Refs.~\onlinecite{Prem_Haad_Nandkishore_2017,Pretko_2017_finite_temperature,PhysRevX.9.021003,PhysRevX.10.011047, khemani_shattering_2020, Rakovszky_Sala_Verresen_Knap_Pollmann2020}). 

An important test case for fracton systems is the limit of random dynamics (infinite temperature), in which one considers the equal-weight statistical average of all possible unitary evolutions. In a typical quantum many-body system, such random dynamics produces, in the long-time average, a thermal ensemble of all states consistent with the conserved quantities (the ``symmetry sector'') \cite{Deutsch_1991,Srednicki_1994,Rigol2008}. But in fracton systems the restricted mobility can lead to a ``fragmentation'' of the symmetry sector into disconnected subsectors that are not mutually accessible via unitary evolution \cite{PhysRevX.10.011047, khemani_shattering_2020}. Such fragmentation precludes thermalization when it occurs in a ``strong'' way, such that all subsectors occupy a small volume of the relevant Hilbert space. The question of which conditions produce or preclude thermalization in fracton systems continues to attract significant interest.

\begin{figure}[htb]
    \centering
    \includegraphics[width=1.0\columnwidth]{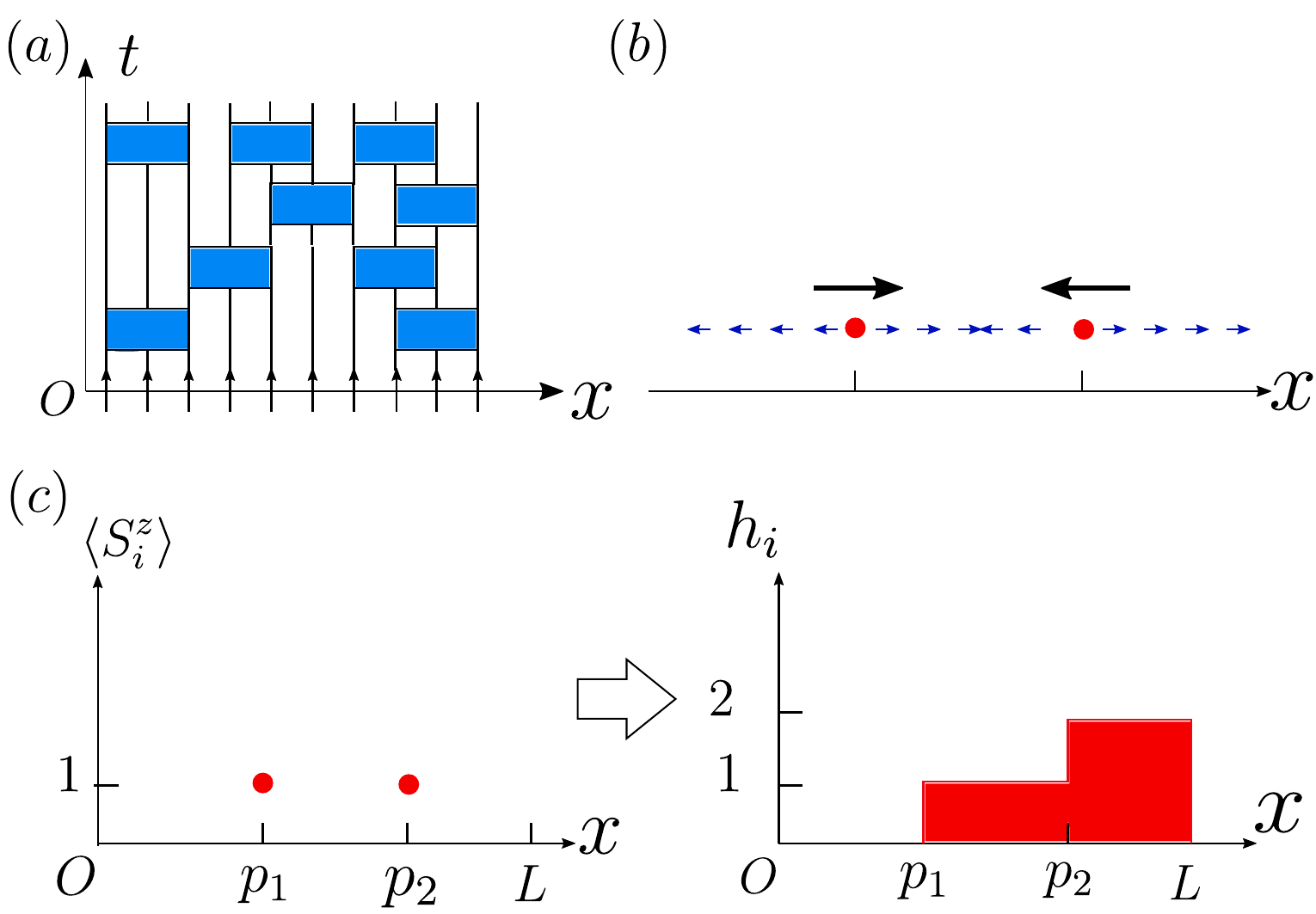}
    \caption{(a) An example of a random unitary circuit acting on a spin-1 chain with three-site gates. Each block represents a random unitary operator that conserves charge and dipole moment. (b) An illustration of the attraction between fractons (red dots). Fractons move in a given direction by emitting dipoles in the opposite direction, or absorbing dipoles from the same direction. In a system with three-site gates, opposite-facing dipoles cannot pass through each other, and this causes inward-facing dipoles to return more quickly to the fracton that emitted them, biasing the motion of fractons toward each other. (c) An illustration of the mapping between fracton charge (left) and the corresponding height field (right).}
    \label{fig0}
\end{figure}

A useful method for describing random dynamics is using the random unitary circuit \cite{Keyserlingk_Sondhi_2018,Khemani_Vishwanath_Huse_2018,Nahum_Ruhman_Vijay_Haah_2017,Nahum_Ruhman_Jonathan_Huse_2018,Nahum_Vijay_Haah_2018, chan_chalker_chaos_2018, chan_spectral_2018}, as depicted in Fig.~\ref{fig0}~(a), in which the system evolves via discrete gates, each of which is chosen randomly among the set of all possible operators that conserve charge and dipole moment. More precisely, one can consider a spin-1 chain, such that the ``charge'' at a given site $i$ corresponds to the expectation value of the operator $S_i^z$, which has eigenstates $-1, 0$, or $+1$. In this description the ``charge density'' at site $i$ is given by $\langle S_i^z \rangle$ (where $\langle ... \rangle$ denotes the quantum-mechanical expectation value) and the total dipole moment of the system $P = \sum_i x_i \langle S_i^z \rangle$, where $x_i$ is the position of the $i$th site.
Unfortunately, a direct simulation of the quantum evolution of a large system is difficult due to the exponentially large Hilbert space. However, a recently-proposed protocol called automaton dynamics (AD) \cite{PhysRevB.100.214301, Alba_Dubail_Medenjak_2019, Iaconis_automata_2021, Gopalakrishnan_automata_2018} overcomes this limitation by focusing on operators that take any product state to another product state (multiplied by an overall phase). Such dynamics can be sampled using classical Monte Carlo, enabling the study of large system sizes and long times.

Previous work has shown that for random unitary dynamics with gate size larger than three, the fracton system is thermalized after a long enough time \cite{PhysRevX.10.011047,PhysRevE.103.022142, PhysRevLett.125.245303, PhysRevB.101.214205, Moudgalya_Prem_Huse_Chan2021}. Under such thermalizing dynamics, the expectation value of any observable (such as the charge density) evolves to that of a thermal ensemble consistent with the fixed charge and dipole moment.  The approach to thermalization is described by a universal subdiffusive dynamics, 
for which a wave vector $q$ is associated with a relaxation time scale $\tau \propto 1/q^4$ \cite{PhysRevResearch.2.033124,PhysRevE.103.022142, PhysRevLett.125.245303, PhysRevB.101.214205, Moudgalya_Prem_Huse_Chan2021} 
(as compared to Brownian diffusion, which has $\tau \propto 1/q^2$).  On the other hand, in a circuit with three-site gates, the system fails to thermalize even at infinite time \cite{PhysRevX.10.011047,PhysRevX.9.021003}. That is, the system retains a memory of the initial state even after infinite time.

In this paper we examine such non-thermalizing dynamics, focusing particularly on unitary circuits with three-site gates. We restrict our attention to the evolution of a spin-1 chain from initial states with a small number of fracton charges.
We show that both the final state and the time scale associated with approaching the final state are governed by very different rules than in the thermalizing case. One of our most striking results (first hinted at by numerical results presented in Ref.~\onlinecite{PhysRevX.9.021003}) is an effect reminiscent of the Casimir effect, in which random fluctuations of the ``dipole field'' produce an effective attraction between two initially separated fracton charges. 

One could anticipate this attraction, in a qualitative sense, by imagining an initial state with two fractons in an otherwise empty system [as in Fig.\ \ref{fig1}(c)]. Each fracton can move (say, to the right) only by emitting a left-facing dipole to its left or absorbing a right-facing dipole from its right. In a common heuristic description (as, e.g., in Ref.~\onlinecite{PhysRevX.9.021003}), one imagines a dipole as a separate kind of particle, which diffuses freely through the system until it encounters a fracton and is absorbed. Importantly, however, under three-site gate dynamics two opposite-facing dipoles cannot pass through each other, and consequently the two fractons effectively reflect each other's emitted dipoles. This reflection causes the dipoles emitted inward to return more quickly than the dipoles emitted outward, leading to a bias in the random motion of the fractons that pulls them together [as depicted in Fig.\ \ref{fig1}(b)]. 
Below we make this heuristic picture more precise, and show how the attraction and its associated time scale can be described using an exact mapping to a simple classical problem involving the sliding of ``blocks'' in an area-preserving height field [depicted in Fig.\ \ref{fig1}(c)].

We emphasize that the attraction between fractons that we discuss in this paper is qualitatively different, and of different origin, than the ``emergent gravity'' between fractons that is discussed in Refs.~\cite{pretko_gravity_2017, prem_phases_2018, pretko_gauge_principle_2018}. We are considering \emph{random} dynamics, which has no associated Hamiltonian or conserved energy, while these previous references discuss systems characterized by a well-defined Hamiltonian. The major point of our paper is to show that even completely random, non-Hamiltonian dynamics produces an effective attraction between fractons, 
but it requires the fragmentation of the Hilbert space that is associated with limited operator size.

The reminder of this paper is organized as follows. We begin in Sec.~\ref{sec:four-size} by considering a random circuit with four-site gates, and showing that the dynamics leads to a thermalized state with a fracton charge density profile that is consistent with a simple maximum-entropy derivation.
Section \ref{sec:three-size} considers three-site gate dynamics, and presents numeric results for the evolution of initial states having either one or two fractons. We present results for the final state and the time scale associated with its approach.  In Sec.~\ref{sec:exact}, we propose an exact mapping to a classical problem of randomly sliding blocks, and we use it to use it to explain our numeric results. We conclude in Sec.~\ref{sec:concl} with a summary and discussion of potential future work.

\section{Simulation method and thermalization with four-site gates}
\label{sec:four-size}

The system we consider is a one-dimensional spin-1 chain with $L$ sites. The basis states of the system can be written as strings of $S^z$ values, with each character in the string denoting the $z$-component of spin at the corresponding site, which takes one of the values $+$, $0$, or $-$.  The total fracton charge is defined by $\qtot = \sum_i\langle S_i^z \rangle$, and the total dipole moment is $\ptot = \sum_i x_i \langle S^z_i\rangle$, where $x_i$ is the coordinate of site $i$. We define our units and coordinate system such that $x_i = i$, with the index $i$ running from $1$ to $L$. 
Each gate in the quantum circuit conserves both the charge and the dipole moment, so that both quantities are constant over time.

Recent work has shown that, under the dynamics of a random unitary circuit with four-site gates, the spin-1 chain thermalizes after a long enough time \cite{PhysRevX.10.011047,PhysRevE.103.022142, PhysRevLett.125.245303, PhysRevB.101.214205,Moudgalya_Prem_Huse_Chan2021}. Consequently, the entropy of a given observable is maximized within the symmetry sector. In fact, even under four-site gate dynamics, a given symmetry sector is fragmented into exponentially many subsectors \cite{PhysRevX.10.011047, khemani_shattering_2020}.  But in the thermodynamic limit one of these subsectors becomes dominant, occupying nearly all of the volume of the Hilbert space of the symmetry sector. This ``weak fragmentation'' allows the dynamics to recover its ergodicity. 

In this section we explicitly check the ergodicity of four-site gate dynamics, and we derive a simple relation for the charge density profile $\langle S^z_i(x_i) \rangle$ based on a maximum-entropy argument. We confirm this theoretical result using two different, complementary numerical simulations. First, we directly construct the maximum entropy result by summing over all basis states within the symmetry sector -- we refer to this procedure as the ``maximum-entropy simulation''. Second, we simulate the four-site dynamics using the AD method. We find that the corresponding numerical results for $\langle S^z_i\rangle$ agree both with each other and with the theory.

The details of our AD simulation method are as follows. We start with a well-defined initial state $\ket{\psi(0)}$, which is a product state and represents a specific configuration of charge. We then apply a random ordering of unitary gates on the initial state, with each gate operating on a randomly chosen set of three nearest-neighboring spins. Each gate is chosen randomly from the set of operations that conserve both the charge and dipole moment (the full set of these operations is enumerated below for three-site gate dynamics, and for four-site dynamics it is summarized in Table I of Ref.~\onlinecite{PhysRevX.9.021003}). Under AD, the resulting state remains a product state after the operation of the gate \cite{PhysRevB.100.214301, Iaconis_automata_2021}. A single time step is defined such that during one time step each spin has, on average, been affected by one gate (i.e., $L/n$ gates constitute a time step for dynamics with $n$-site gates). 
A single realization $j$ of this protocol produces a product state $\ket{\psi(t)}_j$, where $t$ denotes the number of time steps. 
To compute the expectation value of some physical operator $\hat{A}$, we average the value of $\bra{\psi(t)}_j\hat{A}\ket{\psi(t)}_j$ over many independent realizations $j$.

\begin{figure}
    \centering
    \includegraphics[width=1.0\columnwidth]{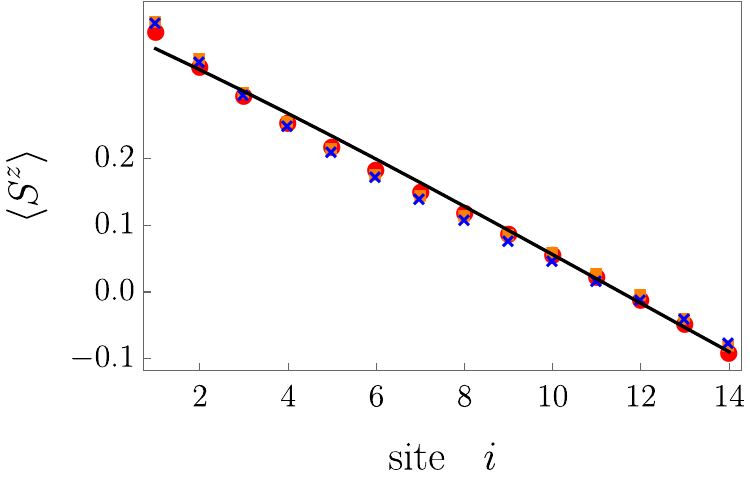}
    \caption{The average fracton charge density $\langle S_i^z \rangle$ as a function of the site index $i$ for a thermalizing system with total charge $\qtot = 2$, dipole moment $\ptot = 7$, and size $L = 14$. Red circles show the results from the maximum-entropy simulation. Orange squares and blue crosses show the results from AD simulations, starting from an initial state with only two fractons (orange squares: initial positions $i = 2, 5$; blue crosses: initial positions $i = 3, 4$). The black curve shows the theoretical result of Eq.~(\ref{eq:Szlinear}). }
    \label{fig1}
\end{figure}

Figure \ref{fig1} shows the value of $\langle S^z_i\rangle$ as a function of the site index $i$ for an example system with size $L = 14$, total charge $\qtot=2$, and total dipole moment $\ptot=7$. The red circles correspond to the maximum-entropy simulation, while orange squares and blue crosses are taken from the AD simulation with different initial states. For the AD simulation, the data corresponds to the state of the system after 45000 time steps and is averaged over 5000 independent realizations. The initial state was chosen to have $S^z_j = 1$ at two specific sites $j$, with all other sites $i$ having $S^z_i = 0$. From Fig.~\ref{fig1}, we see that choosing $j = 3, 4$ and $j = 2, 5$ yield numerically indistinguishable results. The close equivalence of the simulation results implies that the system is thermalized after a sufficiently long time, and all information from the initial state is lost.

We can understand the curve $\langle S^z(x_i) \rangle$ quantitatively using a simple entropy maximization argument. Consider that each site $i$ has some probability $p_i(s)$ of having the $S^z$ value $s = +, 0$, or $-$, with $\sum_{s} p_i(s) = 1$. In the limit where there are many basis states in a given symmetry sector, we can take the probabilities $p_i(s)$ to be independent for different sites $i$, so that the probability of a given string $\{s_1, s_2, ... , s_L\}$ is given by $\prod_i \,p_i(s_i)$. The corresponding Shannon entropy of the probability distribution is
\be
H = -\sum_{i,s}p_i(s)\ln p_{i}(s).
\ee
For a given symmetry sector, there are two constraints on the probabilities $p_i(s_i)$:
\begin{eqnarray}
\qtot =\sum_{i,s} s p_i(s),\\
\ptot =\sum_{i,s} x_i \times s p_i(s).
\end{eqnarray}

We can extremize the value of $H$ using the method of Lagrange multipliers, which gives
\be
p_i(s)=\frac{\text{e}^{s\left(\lambda_Q + x_i \lambda_P\right)}}{1+2\cosh\left(\lambda_Q+x_i\lambda_P\right)}.
\ee
Here, $\lambda_{Q}$ and $\lambda_{P}$ are Lagrange multipliers whose values correspond to the solutions of the nonlinear equations
\begin{eqnarray}
\sum_i\frac{2\sinh{\left(\lambda_Q+x_i\lambda_P\right)}}{1+2\cosh{\left(\lambda_Q+x_i\lambda_P\right)}}=\qtot,\label{eqn:5}\\
\sum_i\frac{2 x_i \sinh{\left(\lambda_Q+x_i\lambda_P\right)}}{1+2\cosh{\left(\lambda_Q+x_i\lambda_P\right)}}=\ptot.\label{eqn:6}
\end{eqnarray}

In the limit $|\qtot| \ll L$ and $|\ptot| \ll L^2$, one can linearize these equations to arrive at
\begin{eqnarray}
&\lambda_{Q}& \simeq \frac{6L\qtot+3\qtot-9\ptot}{L(L-1)}\label{eqn:7}\\ &\lambda_P &\simeq \frac{-9(L\qtot+\qtot-2\ptot)}{L(L^2-1)}.
\label{eqn:8}
\end{eqnarray}
In the same limit, the corresponding value of $\langle S_i^z \rangle = p_i(+) - p_i(-)$ is given by the simple linear equation
\be 
\langle S_i^z \rangle \simeq \frac{2}{3} \left( \lambda_Q + x_i \lambda_P \right).
\label{eq:Szlinear}
\ee 
This analytical result is plotted as the black line in Fig.~\ref{fig1} using the parameters from the simulations, $\qtot = 2$ and $\ptot = 7$.

\section{Nonthermalizing dynamics with three-site gates}
\label{sec:three-size}

As mentioned in the introduction, the dynamics of a random circuit with three-site gates is very different from that of a circuit with four-site gates. In the three-site gate case, the system fails to thermalize due to strong fragmentation of the symmetry sector \cite{PhysRevX.10.011047}, and the long-time behavior of observables cannot be anticipated using maximum-entropy arguments. Instead, different initial states within the same symmetry sector may evolve to produce qualitatively different values of a given observable. In this section we consider the dynamics of the charge density profile $\langle S_i^z \rangle$ under three-site gate dynamics, starting from initial states with only one or two fractons. We present numerical results here, and characterize the effective attraction between fractons and between fractons and the boundary of the system. Section \ref{sec:exact} provides an analytical description of these phenomenona via an exact mapping.


Under three-site gate dynamics, the set of operations is very limited by the symmetry constraints. Table \ref{tab0} gives the full set. Only eight of the $3^3$ possible three-site basis strings admit a nontrivial operation that conserves charge and dipole moment. In the heuristic language of the introduction, the top line of Table \ref{tab0} may be viewed as single fracton hopping left (or right) while emitting a right-facing (left-facing) dipole. The bottom two lines can be viewed as the free translation of a dipole.


\begin{table}[htb]
    \centering
    \begin{tabular}{ccc}
    
      $0+0$ &$\longleftrightarrow$& $+-+$\\
    
     $0-0$&$\longleftrightarrow$& $-+-$ \\
     
     $+-0$&$\longleftrightarrow$& $0+-$\\
     
     $-+0$&$\longleftrightarrow$& $0-+$\\
    
    \end{tabular}
    \caption{The set of all allowed three site gates that conserve charge and dipole moment.}
    \label{tab0}
\end{table}

\subsection{Single fracton initial state}
\begin{figure}
    \centering
    \includegraphics[width=1.0\columnwidth]{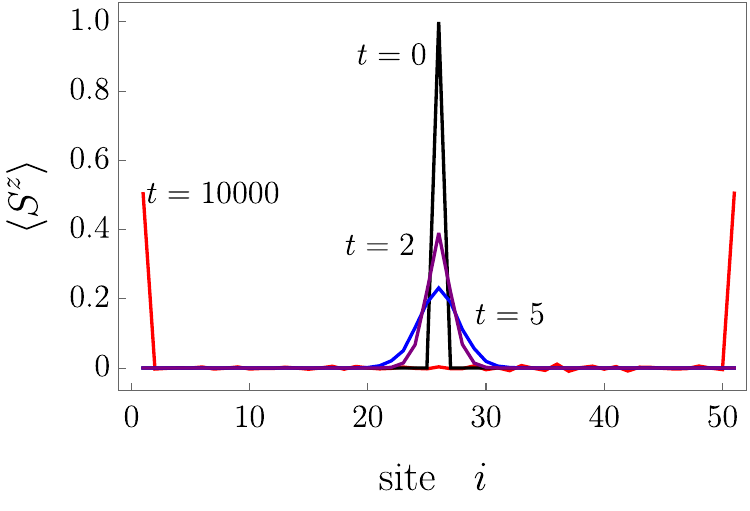}
    \caption{The evolution of the charge density starting from a single fracton peak. The data corresponds to AD simulations with three-site gates for a system with size $L=51$. At $t=0$, we have a single fracton peak in the middle of the system. As time progresses, the peak gradually spreads outward, before eventually accumulating at the system boundaries.}
    \label{fig2}
\end{figure}

A simple starting point for examining the dynamics of nonthermalizing states is to consider initial states for which there is only a single fracton charge (which, for definiteness, we take to be positive) located at site index $i_0$ (position $x_0$).  In other words, the initial state is $\ket{\psi(0)}=\ket{+}_{i_0}\otimes \prod_{i\neq i_0}\ket{0}_i$. 
The AD simulation method \cite{PhysRevE.103.022142} allows us to extract the charge density profile $\langle S_i^z \rangle$ as a function of time. 
The result is shown in Fig.~\ref{fig2} for the case where $i_0$ corresponds to the center of the system. As time increases, the delta-function peak of charge spreads outward, as one would expect. However, rather than spreading to uniformly fill the system, as one would naively anticipate from the maximum entropy state, at late times the fracton charge accumulates at the boundaries of the system. After a very long time, the fracton charge is completely concentrated into two delta-function peaks, one at either boundary. The weight of the two delta-function peaks is such that the charge and dipole moment from the initial state are preserved [i.e., the right edge of the system has charge $(x_0 - 1)/(L-1)$ and the left edge has charge $(L - x_0)/(L-1)$].  

\label{sec:single}

\begin{figure}
    \centering
    \includegraphics[width=1.0\columnwidth]{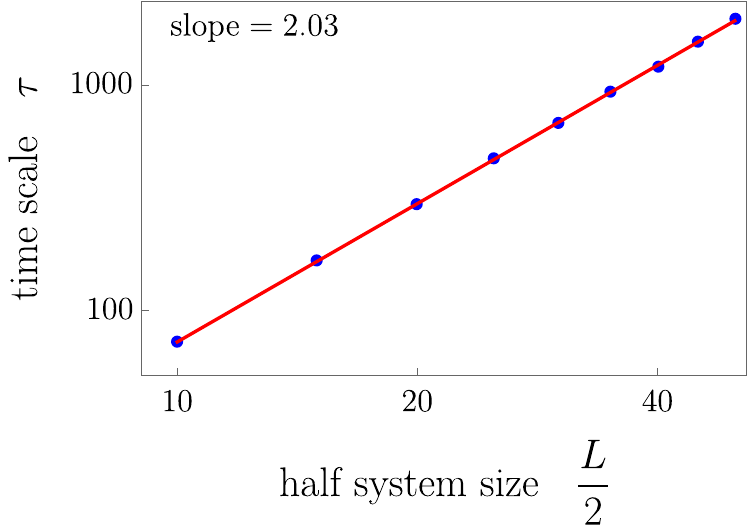}
    \caption{The time scale $\tau$ associated with the approach to the final state, plotted as a function of the half-system size $L/2$, given an initial state with a single fracton at the center. The straight-line fit (in log-log scale) with slope close to $2$ implies a diffusive dynamics, as opposed to the subdiffusive dynamics that arises in thermalizing fracton systems. The standard error of the numeric points (dots) is smaller than the symbol size.}
    \label{fig3}
\end{figure}

In order to characterize the time scale associated with the approach to the final state, we define the quantity
\be
r(t) =\sqrt{ \sum_i \langle S^z_i\rangle(t)\times (x_i-x_0)^2},
\ee
which describes the effective width of the fracton peak as a function of time.
We consider the time scale $\tau$ such that $r(\tau) = L/4$, given an initial state with $x_0 = L/2$. 
Figure \ref{fig3} shows the value of this time scale as a function of system size, along with a power law fit. The fitted exponent ($\tau \propto L^{2.03}$) suggests a quadratic dependence of the dynamical time scale on the system size, which stands in contrast to the universal dynamics $\tau\sim L^4$ of thermalizing fracton systems \cite{PhysRevResearch.2.033124,PhysRevE.103.022142, PhysRevLett.125.245303, PhysRevB.101.214205,Moudgalya_Prem_Huse_Chan2021} . 

\subsection{Attraction between fractons}
\label{sec:double}

We now consider the case where the initial state consists of two fractons: $\ket{\psi(0)}=\ket{+}_{i_1}\otimes\ket{+}_{i_2}\otimes\prod_{i \neq i_1,i_2}\ket{0}_i$, where $i_1$ and $i_2$ are the indices of the two initial fracton positions.

\begin{figure}
    \centering
    \includegraphics[width=1.0\columnwidth]{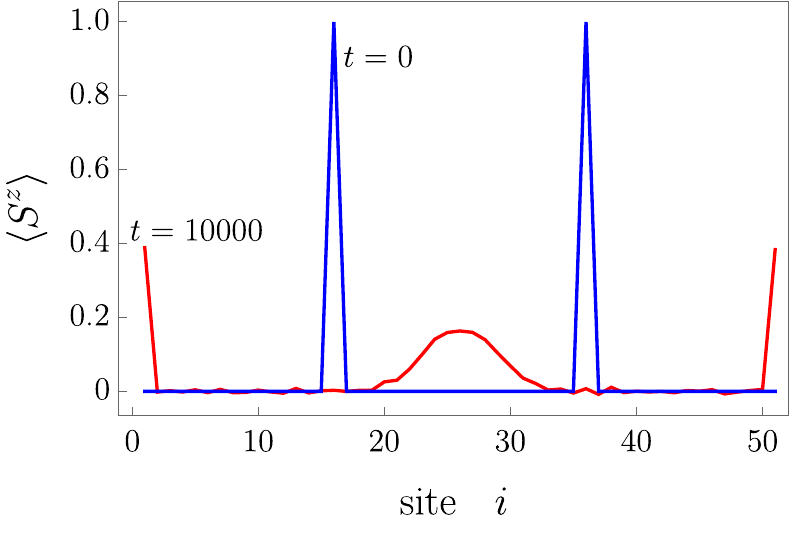}
    \caption{The initial and final charge density profile $\langle S^z_i \rangle$ for an initial state with two fractons that evolves under three-site-gate dynamics. The effective attraction between fractons is manifest in a persistent peak of charge density that forms at the midpoint between the initial fracton positions after a long time (red curve). In this example the system size $L = 51$ and the two initial fracton positions are $i_1=16$ and $i_2=36$ respectively (blue curve).}
    \label{fig4}
\end{figure}

Figure \ref{fig4} shows numerical results for the initial-time and late-time charge density profiles for an initial state with $i_1=16$ and $i_2=36$ and system size $L = 51$, as given by an AD simulation. As in the single-fracton case, the boundary of the system exhibits delta-function peaks of charge density. More striking, however, is the appearance of a persistent, localized peak of charge density at the midpoint between the two initial fracton positions. This peak persists even in the limit of infinite time, and has a width significantly smaller than the distance between the two initial fracton positions.

Since the dynamics in our system is explicitly reversible, this effective attraction between the two fracton peaks can be said to have an entropic origin. That is, within the subsector that includes the initial state, basis states tend to have positive charge near the midpoint between the two initial fracton peaks. An intuitive rationale for this peak is similar to the one given in the introduction: a single fracton moves by emitting dipoles, but under three-site gate dynamics these dipoles cannot pass through each other or through another fracton charge. In this sense each fracton serves like a ``boundary condition'' for the dipoles emitted by other fractons, in loose analogy with the fluctuations of the electromagnetic field that drive the Casimir effect, and the total configurational entropy of the emitted dipoles is maximized when the two fractons are pushed together. 
Below we provide a more precise way of describing this effect, which does not rely on an artificial separation of the system's $+$ and $-$ charges into ``fractons'' and ``dipoles.''


In order to study the time scale associated with the attraction between the two fracton peaks, we define the quantity $\langle r^{\prime}\rangle$
\be
r^{\prime}(t) =
\sum_{\substack{i>L/2, \\ i\ne L}} \langle S^z_i\rangle(t)\times x_i - \sum_{\substack{i<L/2, \\ i\ne 1}} \langle S^z_i\rangle(t)\times x_i .
\ee
Starting from an initial state where the two fracton peaks are equidistant from the center of the system, this quantity represents the average distance between charge at $x<L/2$ and $x>L/2$. In the definition of $r^\prime (t)$ we neglect the charge peaks at the two boundaries. We define a time scale $\tau$ based on the time at which the two fracton peaks approach each other to half their initial separation: $r^{\prime}(\tau) = \Delta /2$, where $\Delta$ is the initial distance between the two fractons. 

\begin{figure}
    \centering
    \includegraphics[width=1.0\columnwidth]{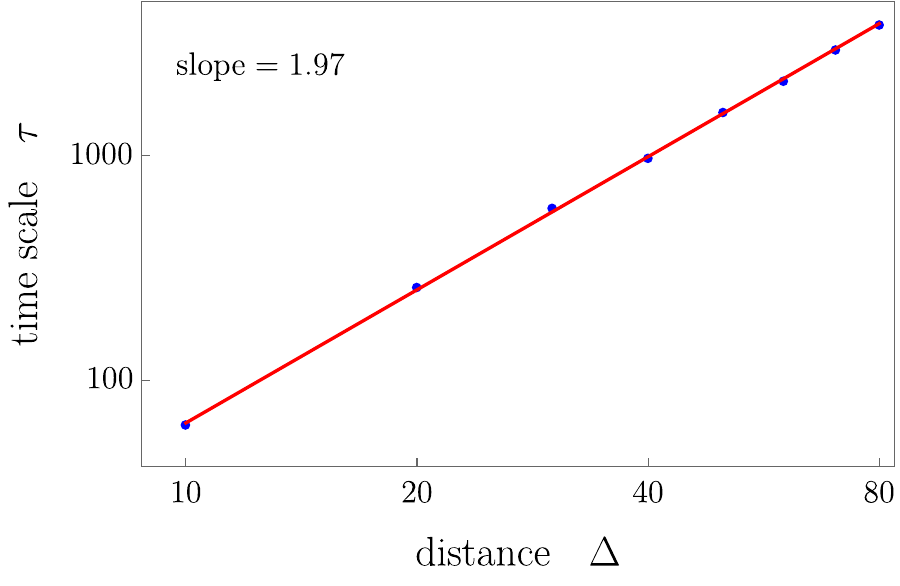}
    \caption{The time scale $\tau$ associated with attraction between two fractons under three-site-gate dynamics is plotted as a function of their initial separation $\Delta$. Blue points correspond to AD simulations with fixed ratio $\Delta/L = 1/2$. The red curve is a power-law best fit, which gives an exponent $\tau \propto \Delta^{1.97}$. The error bars for each point are smaller than the symbol size.}
    \label{fig5}
\end{figure}

Figure~\ref{fig5} shows the behavior of this time scale as a function of the distance $\Delta$, with the ratio $\Delta/L$ kept constant. As in the single fracton case, the time scale $\tau$ increases quadratically with the length scale $\Delta$, suggesting that the dynamics is governed by Brownian diffusion rather than by subdiffusion. 

\section{An exact mapping solution}
\label{sec:exact}

In the previous section we showed, numerically, two dynamical phenomena associated with non-thermalizing dynamics that are starkly different from the thermalizing case. First, we showed that fractons exhibit an effective attraction both to each other and to the boundaries of the system. Second, we showed that the time scale associated with the evolution of the fracton charge density is diffusive, $\tau \propto L^2$, rather than subdiffusive ($\tau \propto L^4$) as in the thermalizing case. In this section we provide a derivation for these behaviors by mapping to a simple problem of classical statistical mechanics.

The key idea is that the fracton charge as a function of position can be mapped to a height field, which evolves according to simple dynamical rules. For a charge distribution $\langle S^z_i \rangle(t)$, we define the ``height''
\be
h_i(t)=\sum_{j\leq i} \langle S^z_i \rangle(t), \quad 0 \le i \le L,
\ee
so that $h_0 = 0$ and $h_L$ is equal to the total charge $\qtot$.
In this mapping, the charge at site $i$ corresponds to the discrete derivative 
\be 
\langle S^z_i \rangle = h_i - h_{i-1}.
\label{eq:hderiv}
\ee 
Importantly, the total area under the height field
\be
\sum_{i=1}^{L-1}h_i(t) = L\qtot-\ptot
\ee
is constant as a function of time. Thus, the fracton dynamics corresponds to area-preserving deformations of the height field $h_i(t)$. 

We note that the same height field mapping is used in Ref.\ \onlinecite{Moudgalya_Prem_Huse_Chan2021} to study thermalizing dynamics, and to derive a generalized hydrodynamic relation.
The entropy of height fields with fixed area has also been studied in the mathematical physics literature (e.g., Refs.~\onlinecite{dobrushin1996fluctuations, perfilev_excursion_2018}). These previous results, however, correspond to the ergodic situation, where the full set of height fields (with fixed area and fixed end points) can be explored. As we now show, under three-site-gate dynamics only a restricted set of height fields is explored.
We mention also that Ref.~\onlinecite{Rakovszky_Sala_Verresen_Knap_Pollmann2020} uses a different, but equivalent classical mapping to label the subsectors of three-site-gate dynamics.

In the language of the height field, the three-site gate operations listed in Table \ref{tab0} correspond to simple operations in which a single ``block'' with unit height and width is slid either to the right or left. This correspondence is outlined in Table \ref{tab1}. 

\begin{table}[]
    \centering
    \begin{tabular}{|c@{\hskip -0.2in}c@{\hskip -0.2in}c||c|}
        \hline
       & \text{fracton charge} && \text{height field} \\
        \hline
        &$+$&&
        \begin{adjustbox}{margin=0mm 1mm 0mm 1mm}
        \begin{tikzpicture}[baseline=-2]
       \draw[thick] (-2.2,-0.2)--(-2.0,-0.2)--(-2.0,0.2)--(-1.8,0.2);
        \end{tikzpicture}
        \end{adjustbox}\\
        \hline
         &$0$&&
        \begin{adjustbox}{margin=0mm 1mm 0mm 1mm}
        \begin{tikzpicture}[baseline=-2]
       \draw[thick] (-2.2,0)--(-2.0,0)--(-2.0,0)--(-1.8,0);
        \end{tikzpicture}
        \end{adjustbox}\\
        \hline
         &$-$&&
        \begin{adjustbox}{margin=0mm 1mm 0mm 1mm}
        \begin{tikzpicture}[baseline=-2]
       \draw[thick] (-2.2,0.2)--(-2.0,0.2)--(-2.0,-0.2)--(-1.8,-0.2);
        \end{tikzpicture}
        \end{adjustbox}\\
        \hline
       $+-+\,$ &$\longleftrightarrow$ &$\,0\,+\,0\,$ & 
       \begin{adjustbox}{margin=1mm 1mm 0mm 1mm}
       \begin{tikzpicture}[baseline=-2]
      \draw [thick] (-2.2,-0.2)--(-2,-0.2)--(-2,0.2)--(-1.6,0.2)--(-1.6,-0.2)--(-1.2,-0.2)--(-1.2,0.2)--(-1.0,0.2);
       \end{tikzpicture}
       \end{adjustbox}\,\,$\longleftrightarrow$\, \begin{adjustbox}{margin=0mm 1mm 0mm 1mm}
       \begin{tikzpicture}[baseline=-2]
       \draw [thick] (-2.2,-0.2)--(-2,-0.2)--(-1.6,-0.2)--(-1.6,0.2)--(-1.2,0.2)--(-1.0,0.2);
       \end{tikzpicture}
       \end{adjustbox}\\
       \hline
       $\,-+-\,$&$\longleftrightarrow$ &$\,0\,-\,0\,$&\begin{adjustbox}{margin=1mm 1mm 0mm 1mm}
       \begin{tikzpicture}[baseline=-2]
       \draw [thick] (-2.2,0.2)--(-2,0.2)--(-2,-0.2)--(-1.6,-0.2)--(-1.6,0.2)--(-1.2,0.2)--(-1.2,-0.2)--(-1.0,-0.2);
        \end{tikzpicture}
       \end{adjustbox}\,\,$\longleftrightarrow$\, \begin{adjustbox}{margin=0mm 1mm 0mm 1mm}
       \begin{tikzpicture}[baseline=-2]
       \draw [thick] (-2.2,0.2)--(-2,0.2)--(-1.6,0.2)--(-1.6,-0.2)--(-1.2,-0.2)--(-1.0,-0.2);
       \end{tikzpicture}
       \end{adjustbox}\\
       \hline
       $\,+\,-\,0\,\,$&$\longleftrightarrow$& $\,0\,+-\,$&\begin{adjustbox}{margin=1mm 1mm 0mm 1mm}
       \begin{tikzpicture}[baseline=-2]
       \draw [thick] (-2.2,-0.2)--(-2,-0.2)--(-2,0.2)--(-1.6,0.2)--(-1.6,-0.2)--(-1.2,-0.2)--(-1.0,-0.2);
       \end{tikzpicture}
       \end{adjustbox}\,\,$\longleftrightarrow$\, \begin{adjustbox}{margin=0mm 1mm 0mm 1mm}
       \begin{tikzpicture}[baseline=-2]
       \draw [thick] (-2.2,-0.2)--(-2,-0.2)--(-1.6,-0.2)--(-1.6,0.2)--(-1.2,0.2)--(-1.2,-0.2)--(-1.0,-0.2);
       \end{tikzpicture}
       \end{adjustbox}\\
       \hline
       $\,-\,+\,0\,\,$&$\longleftrightarrow$ &$\,0\,-+\,$&\begin{adjustbox}{margin=1mm 1mm 0mm 1mm}
       \begin{tikzpicture}[baseline=-2]
       \draw [thick] (-2.2,0.2)--(-2,0.2)--(-2,-0.2)--(-1.6,-0.2)--(-1.6,0.2)--(-1.2,0.2)--(-1.0,0.2);
       \end{tikzpicture}
       \end{adjustbox}\,\,$\longleftrightarrow$\, \begin{adjustbox}{margin=0mm 1mm 0mm 1mm}
       \begin{tikzpicture}[baseline=-2]
       \draw [thick] (-2.2,0.2)--(-2,0.2)--(-1.6,0.2)--(-1.6,-0.2)--(-1.2,-0.2)--(-1.2,0.2)--(-1.0,0.2);
       \end{tikzpicture}
       \end{adjustbox}\\
       \hline
    \end{tabular}
    \caption{Above: the set of all charges and their corresponding height fields. The site locates at the middle of the height field. Below: the set of all possible three-site gates and their corresponding height fields. The left column shows their effects on the fracton charge. The right column shows the representation of this change in the height field. Each operation can be viewed as a single block sliding either right (the fourth or sixth rows) or left (the fifth and seventh rows).}
    \label{tab1}
\end{table}

In this language, the height field can be viewed as a collection of blocks, stacked on top of each other to form some profile $h_i(t)$. Under the action of the random circuit these blocks are moved around according to the following simple rules:
\begin{itemize}
    \item  A three-site gate corresponds to a single block sliding either to the left or to the right by one unit. Blocks may not slide through each other.
    \item Any move that would cause the height field to change by two units at one position $i$ is prohibited.
\end{itemize}
We now show that this simple description of sliding blocks permits an exact solution for both the single-fracton and two-fracton problems.

\subsection{Single fracton block mapping}
\label{sec:single-block}

\begin{figure}
    \centering
    \includegraphics[width=1.0\columnwidth]{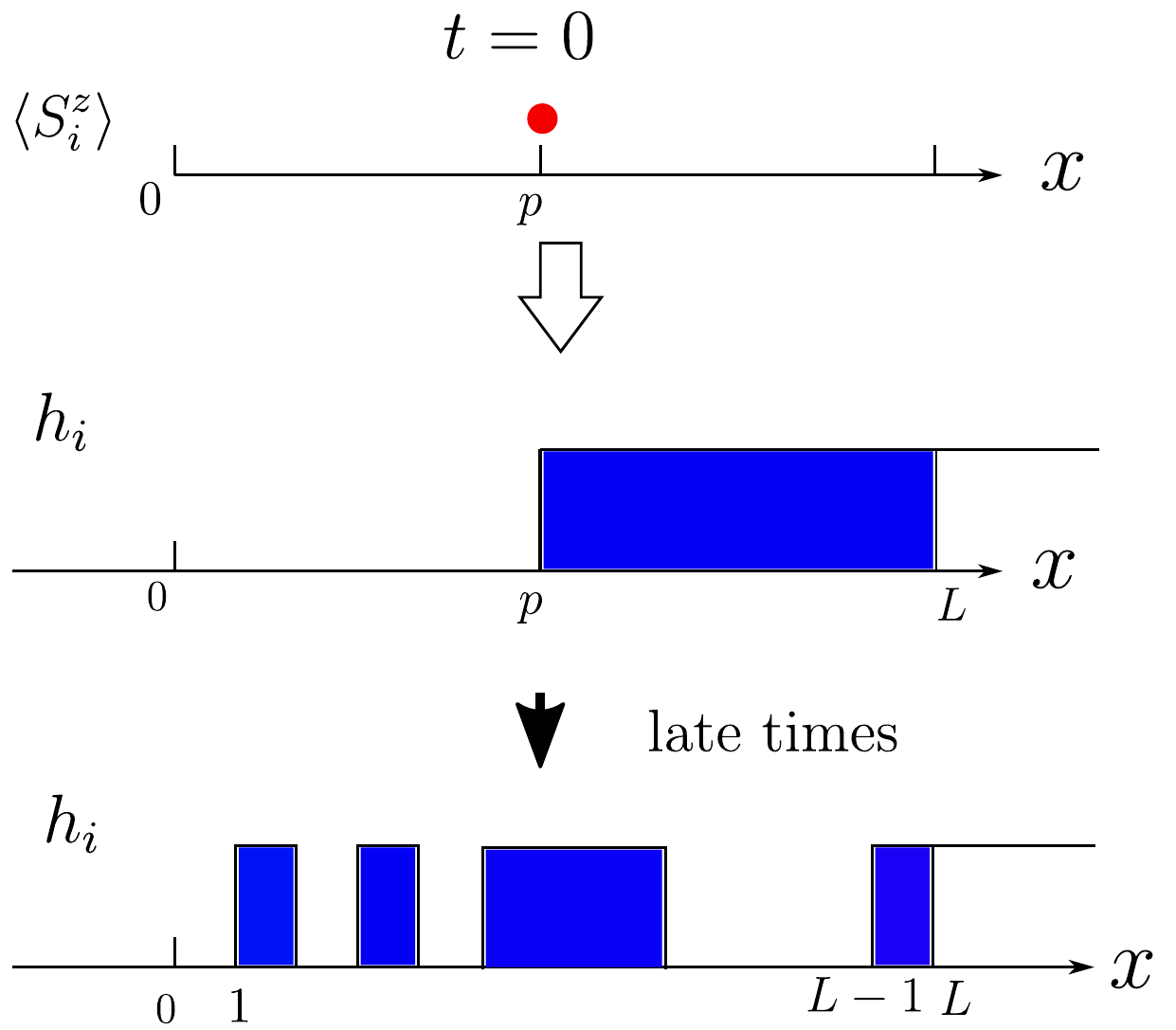}
    \caption{An illustration of the block mapping that describes three-site-gate dynamics for initial states with a single fracton. The top plot shows an initially empty state with only a single fracton charge at position $p$. The white arrow shows the mapping to a collection of $L-p$ contiguous blocks on the right side of the system (middle plot). The height field at positions $i=0$ and $i = L$ is fixed to $h_0=0$ and $h_L=1$, respectively. The black arrow indicates time evolution to a state where the blocks are uniformly scattered across the system (bottom plot).}
    \label{fig6}
\end{figure}

We first consider the case where the initial state contains only a single fracton at position $p$. In this situation the height field at time zero satisfies
\begin{eqnarray}
h_i(0)=\left\{\begin{array}{ll}0,&\quad i<p,\\
1,& \quad i\ge p.
\end{array}\right.
\end{eqnarray}
This mapping is illustrated by Fig.~\ref{fig6}, with the blue area corresponding to $L-p$ contiguous blocks. At the initial time, the only allowed move is for the leftmost block to slide by one unit to the left. Over time, vacancies open up in the chain of blocks, allowing others to move, and the distribution of blocks becomes more uniform. 
This process of diffusion with excluded volume is described mathematically by the so-called simple exclusion process \cite{Komorowski2012}. In the continuous limit ($L, t \gg 1$), the physics is controlled by the usual one-dimensional diffusion equation \cite{Komorowski2012}
\be
\frac{\partial \rho(x,t)}{\partial t}= a\frac{\partial^2\rho(x,t)}{\partial t^2}.
\label{eq:hydro}
\ee
Here $\rho(x,t)$ is the average block density and $a$ (the diffusion constant) is an order-$1$ parameter determined by the details of the circuit. 
Although an exact analytical solution for $\rho(x,t)$ is mathematically complicated due to the boundary conditions in our problem, Eq.~(\ref{eq:hydro}) gives an immediate explanation for the diffusion-like time scale $\tau \propto L^2$ shown in Fig.~\ref{fig4}.

This mapping also gives a simple explanation for the delta-function peaks of charge density that appear at the boundaries of the system after a long time. In the block description, it is clear that after a very long time the block density becomes spatially uniform, ${\rho_i = (L-p)/(L-1)}$. Thus, the height field adopts an average value ${\langle h_i \rangle = (L-p)/(L-1)}$ for all $i$ in the interval $1, ..., L-1$. On the other hand, $h_0 = 0$ and $h_L = 1$ are fixed, so that taking the discrete derivative [Eq.~(\ref{eq:hderiv})] gives 
\be 
  \langle S^z_i \rangle =
  \begin{cases}
        (L-p)/(L-1), & i=1 \\
        0, & 0 < i < L-1 \\
        (p-1)/(L-1), & i=L
  \end{cases}
\ee 
This expression is consistent with the numeric results shown in Fig.~\ref{fig2}, and one can check that it preserves the total charge $\qtot = 1$ and dipole moment $\ptot = p$.

It is worth noting that this simple reasoning can similarly be applied to any initial state containing an alternating pattern of $+$ and $-$ charges (with any arrangement of $0$'s between them). Hamiltonian dynamics of such states was considered in detail in Ref.~\onlinecite{Rakovszky_Sala_Verresen_Knap_Pollmann2020}. In the language of the height field, such states correspond to a height field that remains always between $0$ and $1$ (or between $0$ and $-1$), so that it can be viewed as a single-tiered arrangement of some number of blocks. Thus, the final state similarly exhibits only two delta-function peaks of charge, whose heights can be determined simply by counting the number of blocks.

\subsection{Double fracton block mapping}
\label{sec:double-block}

\begin{figure}
    \centering
    \includegraphics[width=1.0\columnwidth]{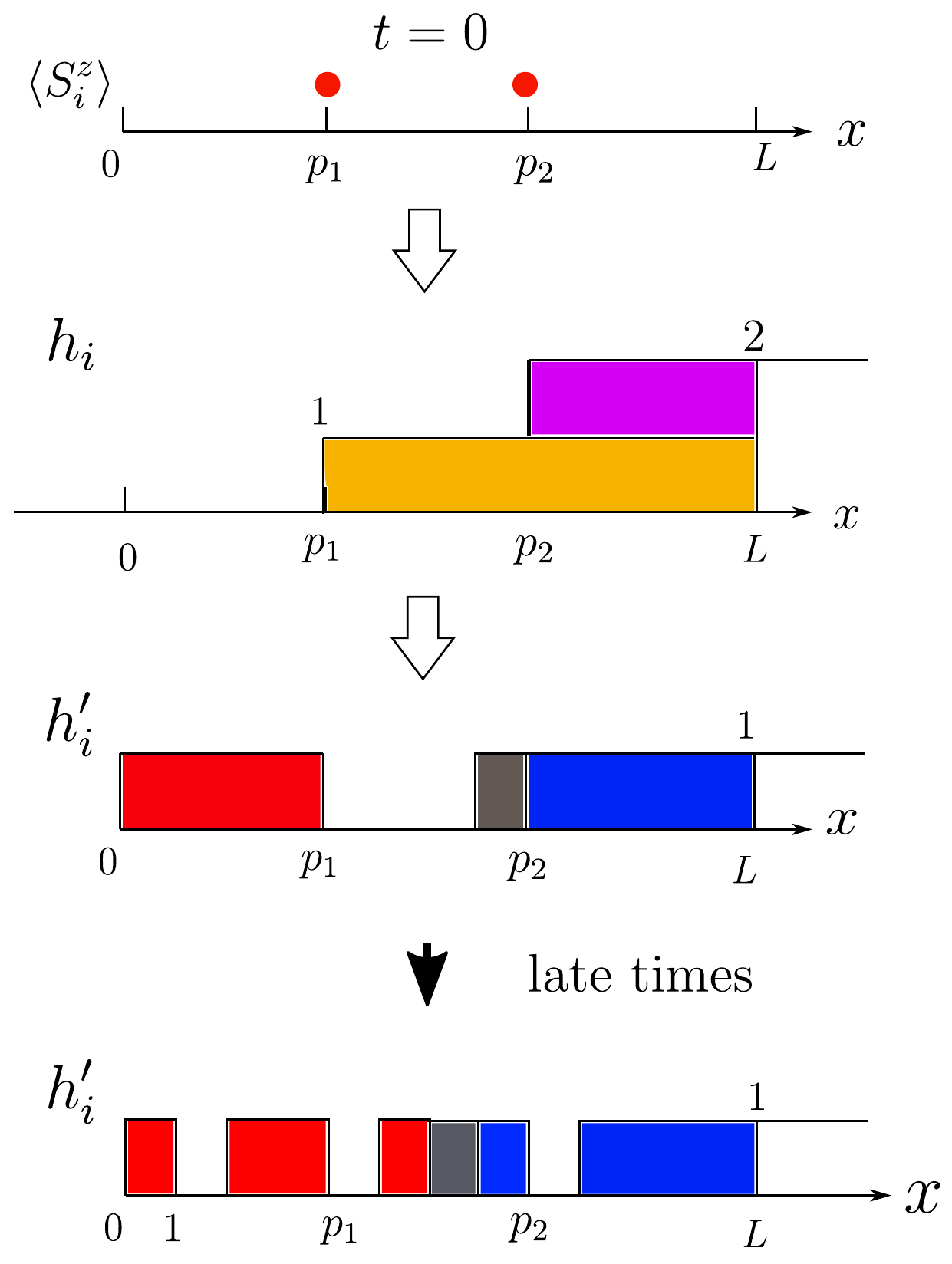}
    \caption{An illustration of the mapping that describes the dynamics of a two-fracton initial state (top plot). At $t=0$, the corresponding height field comprises a layer of blocks with $h = 2$ (purple, second plot) atop a layer of blocks with $h=1$ (orange). A further mapping (third plot) describes site with $h=2$ as particles (blue blocks) and sites with $h = 0$ as holes (red blocks), separated by a movable piston (gray block). The random circuit causes the particles and holes to diffuse inward (bottom plot), filling the empty space between them.}
    \label{fig7}
\end{figure}

A more complicated case is the initial state with two fractons. In this case mapping to the height field gives two layers of blocks, as shown in Fig.~\ref{fig7}. 
In principle, one can directly apply the dynamical rules listed at the beginning of this section to both layers of blocks in order to understand the dynamics. However, a further mapping makes the dynamics easier to understand. We introduce the modified height field
\be
h^{\prime}_i=h_i-1,
\ee
which is defined such that the reference state ($h^\prime = 0$) is a filled first layer of blocks ($h = 1$). We can then discuss two different kinds of excitations in the modified height field: ``particles'' (denoted by blue blocks in Fig.~\ref{fig7}) that correspond to $h^\prime = 1$ and ``holes'' (red blocks in Fig.~\ref{fig7}) that correspond to $h^\prime = -1$. The dynamical evolution of the system consists of the particles and holes diffusing to fill the empty middle region of the system, like two gases expanding into a region of vacuum between them. One subtlety is that, since the height field may not change by two units at one site, the leftmost particle cannot be immediately adjacent to the rightmost hole. Instead, these two must be separated by at least one space, which we describe by a freely-sliding ``piston'' (gray block in Fig.~\ref{fig7}) that separates the two ``gases''.

Since both gases expand outward via the diffusion equation, the observed time scale $\tau \propto \Delta^2$ [see Fig.~\ref{fig5}] is natural. The final state is also easy to understand: it corresponds to a uniform spatial distribution of blocks (red and blue combined). That is, $\rhored(x,t) + \rhoblue(x,t)$ becomes approximately constant as a function of position at late times (with the only caveat being the single piston block). The localized peak in fracton charge $\langle S_i^z \rangle$ is dictated by the statistics of the interface between the two gases. Specifically, since blue particles correspond to positive $h^\prime$ and red holes to negative $h^\prime$, the charge density is given by
\be
\langle S_i^z\rangle(x) = \rhoblue(x_i) - \rhoblue(x_{i-1}) - [\rhored(x_i) - \rhored(x_{i-1})].
\ee
The charge density therefore has a peak at the midpoint between the two gases, where $\rhored$ is declining with increased index $i$ while $\rhoblue$ is increasing. The width of this peak is of order $\sqrt{\Delta}$, which is much smaller than the initial distance $\Delta$ between the two fractons.

\begin{figure}
    \centering
    \includegraphics[width=1.0\columnwidth]{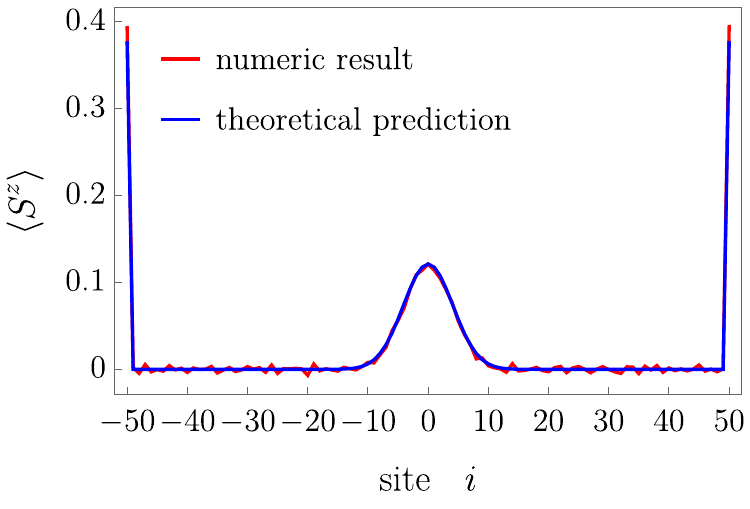}
    \caption{The late-time charge density $\langle S_i^z\rangle$ of a system with two fractons in the initial state, as given by three-site-gate dynamics. In this example the two fractons are initially at positions $i=-20$ and $i=20$, respectively. The red curve shows the result of numerical simulations, and the blue curve is our analytical result.}
    \label{fig8}
\end{figure}

A full derivation of the charge density $\langle S_i^z \rangle$ at late times is presented in Appendix \ref{sec:appendix}. The resulting analytical curve is plotted in Fig.~\ref{fig8} (blue curve), and it very closely matches the numerical result.

\section{Conclusion}
\label{sec:concl}

In this paper we have explored the dynamics of nonthermalizing fracton systems, focusing on the case of a spin-1 chain operated on by a random unitary circuit that conserves charge and dipole moment. In the case where the circuit contains four-site gates, the dynamics is thermalizing. We have shown, in particular, that in this case the distribution of charge density after a long time can be captured by a simple maximum-entropy argument, and there is no need to know the specific initial state. On the other hand, the dynamics with only three-site gates is highly nonthermalizing, such that different initial states within the same symmetry sector evolve to produce strongly different distributions of charge density after a long time. 

Most striking among our results is that initial peaks of fracton charge density exhibit an effective attraction toward each other and toward the boundaries of the system. The tightly-localized peaks of charge density which appear at late times are in strong contrast to the thermalizing case, for which early-time peaks of charge density tend to spread uniformly across the system. This difference arises fundamentally from the strong fragmentation of the Hilbert space that occurs in the nonthermalizing case.

We explain the dynamics in the nonthermalizing case using a simple, exact mapping to a classical statistical mechanics problem of sliding blocks in a height field. The mapping gives a natural explanation for both the appearance of localized charge peaks at infinite time and for the diffusion-like time scale $\tau \propto L^2$ associated with the system's approach to the steady state. 

In principle, the mapping we present gives a natural way to describe the dynamics of arbitrary initial states. So far, however, we have only focused on two simple cases, which correspond to the height field having either one or two tiers. Generalizing our approach to a more generic set of initial states may be a promising direction for future work.

Unfortunately, the height field mapping does not imply an obvious generalization to higher dimensions. But the question of whether a similarly simple statistical mechanics description can capture the dynamics of two-dimensional or higher-dimensional nonthermalizing fracton systems is an interesting one.

\acknowledgments 
The authors are grateful to Xiaoyu Feng, Andrey Gromov, Jason Iaconis, Adam Nahum, Rahul Nandkishore, and Shriya Pai for helpful discussions, and to Pablo Sala and Michael Knap for bringing relevant references to our attention. This work was primarily supported by the Center for Emergent Materials, an NSF-funded MRSEC, under Grant No. DMR-2011876.

\appendix
\section{The exact solution of the final state for two-fracton problem}
\label{sec:appendix}

In this Appendix we present the full derivation of the late-time charge distribution for an initial state with two fractons. As mentioned in Sec.~\ref{sec:double-block}, we map the problem onto the problem of two kinds of blocks, red and blue, diffusing to fill the initially empty space between them. The average densities of the two kinds of blocks are given by $\rhored(x,t)$ and $\rhoblue(x,t)$, respectively.

Separating the two kinds of blocks is a movable `piston', which can be viewed as a special boundary between them. The densities of red and blue blocks can be written as
\begin{eqnarray}
\rhored(x,t)=\sum_{\xi}\mathcal{W}(\xi,t)\rhored(x,t;\xi),\label{eqn:16}\\
\rhoblue(x,t)=\sum_{\xi}\mathcal{W}(\xi,t)\rhoblue(x,t;\xi).\label{eqn:17}
\end{eqnarray}
Here $\xi$ labels the position of the piston and $\rho(x,t;\xi)$ means the average density for those states with the piston located at $\xi$ at time $t$. $\mathcal{W}(\xi,t)$ is a weight function that represents the probability distribution of $\xi$ at time $t$. 

We expect that after the final state is reached, the weight function no longer changes with time and can be written as $\mathcal{W}(\xi)$. Also, since the densities of blocks no longer change,
\begin{eqnarray}
\frac{\partial \rhored}{\partial t}=\sum_{\xi}\mathcal{W}(\xi)\frac{\rhored(x,t;\xi)}{\partial t}=0,\\
\frac{\partial \rhoblue}{\partial t}=\sum_{\xi}\mathcal{W}(\xi)\frac{\rhoblue(x,t;\xi)}{\partial t}=0,
\end{eqnarray}
which lead to a simple solution
\be
\frac{\partial \rhored(x,t;\xi)}{\partial t}=\frac{\rhoblue(x,t;\xi)}{\partial t}=0.
\ee
This equation means that a static solution is reached for red and blue blocks with a specific $\xi$. Recalling that the diffusion equation is satisfied for red and blue blocks respectively, a static solution should be a constant function or a linear function. The latter is forbidden since a nonzero slope means the change of block density on the boundaries. Thus we conclude that for the final state, 
\begin{eqnarray}
&\rhored(x,\xi)=\left\{\begin{array}{cc}
    \frac{L/2-\Delta/2}{L/2+\xi}, &\quad x<\xi, \\
    \\
    0, &\quad x>\xi, 
\end{array}\right.\label{eqn:21}\\
\nonumber\\
&\rhoblue(x,\xi)=\left\{
\begin{array}{cc}
    0, &\quad x<\xi  \\
    \\
    \frac{L/2-\Delta/2}{L/2-\xi}, &\quad x>\xi 
\end{array}\right.\label{eqn:22}
\end{eqnarray}
Here the origin of the $x$ axis is chosen to be the middle of the system. $\Delta$ is the distance between the two initial fracton peaks, and for the initial state, we have chosen the two peaks to be symmetric about the middle of the system. The position positions obeys
\be
-\Delta/2<\xi<\Delta/2.\label{eqn:23}
\ee
This condition is consistent with the conservation of the block area. 

To derive the weight function $\mathcal{W}(\xi)$, we use the fact that in the final state, the rate at which the piston jumps from $\xi$ to $\xi+1$ is equal to the rate at which it jumps from $\xi+1$ to $\xi$. The corresponding equilibrium condition is
\be
\mathcal{P}(\xi\to \xi+1)\mathcal{W}(\xi)=\mathcal{P}(\xi+1\to\xi)\mathcal{W}(\xi+1).\label{eqn:24}
\ee
In other words,
\be
\frac{\mathcal{P}(\xi+1\to\xi)}{\mathcal{P}(\xi\to\xi+1)}=\frac{\mathcal{W}(\xi)}{\mathcal{W}(\xi+1)}.\label{eqn:25}
\ee

In order for the piston to jump left or right, the target site must be empty of blue or red blocks. Since the piston has same probability to jump to left or right,
\begin{eqnarray}
&\mathcal{P}(\xi\to\xi+1)\propto 1-\rhoblue(\xi+1,\xi)\label{eqn:26}\\
&\mathcal{P}(\xi\to \xi-1)\propto 1-\rhored(\xi-1,\xi)\label{eqn:27}
\end{eqnarray}
Plugging Eqs.~(\ref{eqn:21})-(\ref{eqn:22}) and Eqs.~(\ref{eqn:26})-(\ref{eqn:27}) into Eq.~(\ref{eqn:25}), we get
\be
\frac{\mathcal{W}(\xi+1)}{\mathcal{W}(\xi)}=\frac{\Delta/2-\xi}{L/2-\xi}\frac{L/2+\xi}{\Delta/2+\xi}.
\ee
For $\xi\ll \Delta, L$,
\be
\frac{d\mathcal{W}(\xi)}{d\xi}\approx -\frac{4(L-\Delta)\xi}{L\Delta}\mathcal{W}(\xi).
\ee
The solution to this equation is a Gaussian function with width $\frac{1}{2}\sqrt{\frac{ L \Delta}{(L-\Delta)}}\ll \Delta, L$. So the assumption $\xi\ll \Delta, L$ is reasonable when $L\to \infty$ and $\Delta/L$ is a finite number. So the weighting function
\be
\mathcal{W}(\xi) \simeq \frac{\sqrt{2(L-\Delta)}}{\sqrt{\Delta L\pi}}\text{e}^{-\frac{2\xi^2(L-\Delta)}{L\Delta}}.\label{eqn:30}
\ee

Plugging Eqs.~(\ref{eqn:21})-(\ref{eqn:22}) and Eq.~(\ref{eqn:30}) into Eqs.~(\ref{eqn:16})-(\ref{eqn:17}), the densities of red and blue blocks are

\begin{align}
\rhored(x)=&
    \int_{-\Delta/2}^{\Delta/2}\mathcal{W}(\xi)\,\frac{\frac{L}{2}-\Delta/2}{\frac{L}{2}+\xi}\,d\xi, & -\frac{L}{2}<x<-\frac{\Delta}{2},  \nonumber\\
  \rhored(x)  =&\int_x^{\Delta/2}\mathcal{W}(\xi)\,\frac{\frac{L}{2}-\Delta/2}{\frac{L}{2}+\xi}\,d\xi,  & -\frac{\Delta}{2}<x<\frac{\Delta}{2},\\
  \rhored(x)  =& 0,&\frac{\Delta}{2}<x<\frac{L}{2},\nonumber
\end{align}
and
\begin{align}
\rhoblue(x)=&
  0,&-\frac{L}{2}<x<-\frac{\Delta}{2},\nonumber\\
 \rhoblue(x)=&   \int_{-\Delta/2}^{x}\mathcal{W}(\xi)\,\frac{\frac{L}{2}-\Delta/2}{\frac{L}{2}-\xi}\,d\xi, & -\frac{\Delta}{2}<x<\frac{\Delta}{2},\\
 \rhoblue(x)=&   \int_{-\Delta/2}^{\Delta/2}\mathcal{W}(\xi)\,\frac{\frac{L}{2}-\Delta/2}{\frac{L}{2}-\xi}\,d\xi,  & \frac{\Delta}{2}<x<\frac{L}{2}.\nonumber
\end{align}

Using the fact $\langle S^z(x)\rangle \simeq -d\rhored/dx + d\rhoblue/dx$ (here the discrete derivative in the height mapping has been approximated with a continuous derivative), we get the charge distribution
\begin{eqnarray}
\langle S^z(x)\rangle =\left\{
\begin{array}{cc}
   0&-\frac{L}{2}<x<\frac{L}{2}\\
   \mathcal{W}(x)\frac{\left(L/2-\Delta/2\right)L}{L^2/4-x^2}&-\frac{\Delta}{2}<x<\frac{\Delta}{2}\\
    0 & \frac{\Delta}{2}<x<\frac{L}{2}\\
\end{array}
\right.
\label{eq:Sz2final}
\end{eqnarray}
This solution gives a localized peak of charge density in the center of the system in the limit of infinite time, as shown in Fig.~\ref{fig8}. The width of this peak is $\sim \sqrt{\Delta}$, much smaller than the initial fracton separation when $\Delta \gg 1$.

For the charge on the boundaries of the system ($x=\pm L/2$), we point out that two additional conditions,
\begin{eqnarray}
&\rhored=1  &\quad x<-\frac{L}{2},\\
&\rhoblue=1  & \quad x>\frac{L}{2},
\end{eqnarray}
are needed to make the block system consistent with the fracton system. Using these equations, the charge on boundary can be obtained
\be
\langle S^z(x)\rangle=1-\int_{-\Delta/2}^{\Delta/2}\mathcal{W}(\xi)\frac{L/2-\Delta/2}{L/2+\xi}\,d\xi, \quad x=\pm\frac{L}{2}.
\label{eq:S2ends}
\ee
Equation (\ref{eq:Sz2final}), together with the value for the boundaries from Eq.~(\ref{eq:S2ends}), is plotted as the blue line in Fig.~\ref{fig8}.

\bibliography{reference.bib}

\end{document}